\documentclass[12pt,a4paper]{article}
\makeatletter
\def\eqnarray{%
\stepcounter{equation}%
\let\@currentlabel=\theequation
\global\@eqnswtrue
\global\@eqcnt\z@
\tabskip\@centering
\let\\=\@eqncr
$$\halign to \displaywidth\bgroup\@eqnsel\hskip\@centering
$\displaystyle\tabskip\z@{##}$&\global\@eqcnt\@ne
\hfil$\displaystyle{{}##{}}$\hfil
&\global\@eqcnt\tw@$\displaystyle\tabskip\z@{##}$\hfil
\tabskip\@centering&\llap{##}\tabskip\z@\cr}
\makeatother
\begin{document}

\title{\sl How To Treat An N--Level System : A Proposal
\thanks{This is prepared for the coming 11--th Numazu Meeting 
(6--8/March/2003)}
}
\author{
  Kazuyuki FUJII
  \thanks{E-mail address : fujii@yokohama-cu.ac.jp }\\
  Department of Mathematical Sciences\\
  Yokohama City University\\
  Yokohama, 236-0027\\
  Japan
  }
\date{}
\maketitle
%
%
%
%
\begin{abstract}
  In this paper we propose a Hamiltonian of the n--level system 
  by making use of generalized Pauli matrices. 
\end{abstract}


%
%
%
%
An atom has usually many (finite or infinite) energy levels. However, to treat 
infinitely many ones at the same time is not realistic, so we treat 
an atom with finite energy levels. In the following we would like to consider 
an atom with $n$ energy levels interacting with external (periodic) fields 
like $\mbox{cos}(\omega t+\phi)$ (an $n$--level system in our terminology). 

In Quantum Optics it is standard to consider a $2$--level system, see 
for example \cite{AE} and \cite{AC}. The $n$--level system might be considered 
as a composite of $2$--level systems. As for recent developments of the 
$2$--level system including constructions of solutions in the strong 
coupling regime, see \cite{CEC}, \cite{BaWr}, \cite{SGD}, \cite{MFr1}, 
\cite{KF5} and their references. See also \cite{MFr3} as a general 
introduction to two--level system, and \cite{MFr2}, \cite{KF2} 
as some applications to Quantum Optics. 
However we would like to treat an $n$--level system in a direct manner. 
Then we meet some troubles. 
First what is a Hamiltonian describing such a system ?  As far as we know 
such a Hamiltonian has not been known. 
As for some examples of $n$--level system see \cite{KF4}, \cite{ZZH}, 
\cite{EF} and their references. 

In this paper we propose such a Hamiltonian. In the $2$--level system the 
Hamiltonian is written by making use of Pauli matrices 
$\{\sigma_{1},\sigma_{3}\}$ (see the references above), 
so in the $n$--level system the Hamiltonian 
should be given by making use of generalized Pauli matrices 
$\{\Sigma_{1},\Sigma_{3}\}$. We will write down it. 
We believe that it is a natural generalization 

Our real aim of this study is to apply some developments in this paper to 
the theory of qudits in Quantum Computation, see for example 
\cite{KF3} and \cite{KuF}. In it we must first of all construct all unitary 
elements in $U(n)$ (a universality of $1$--qudit).

\par 
We expect that this work will become a starting point toward this direction 
\footnote{This paper may be deeply connected to \cite{ZZH} in Quantum 
Information Theory.}.

\par \vspace{3mm}
Let $\{\sigma_{1}, \sigma_{2}, \sigma_{3}\}$ be Pauli matrices and 
${\bf 1}_{2}$ a unit matrix : 
\begin{equation}
\sigma_{1} = 
\left(
  \begin{array}{cc}
    0& 1 \\
    1& 0
  \end{array}
\right), \quad 
\sigma_{2} = 
\left(
  \begin{array}{cc}
    0& -i \\
    i& 0
  \end{array}
\right), \quad 
\sigma_{3} = 
\left(
  \begin{array}{cc}
    1& 0 \\
    0& -1
  \end{array}
\right), \quad 
{\bf 1}_{2} = 
\left(
  \begin{array}{cc}
    1& 0 \\
    0& 1
  \end{array}
\right), 
\end{equation}
and 
$\sigma_{+} = (1/2)(\sigma_{1}+i\sigma_{1})$, 
$\sigma_{-} = (1/2)(\sigma_{1}-i\sigma_{1})$. 

First list the well--known properties of $\sigma_{1}$ and $\sigma_{3}$ : 
\begin{equation}
\sigma_{1}^{2}=\sigma_{3}^{2}={\bf 1}_{2},\quad 
\sigma_{1}^{\dagger}=\sigma_{1}, \quad 
\sigma_{3}^{\dagger}=\sigma_{3}, \quad 
\sigma_{3}\sigma_{1}=-\sigma_{1}\sigma_{3}=
\mbox{e}^{\pi i}\sigma_{1}\sigma_{3}.
\end{equation}

\par \noindent
Let $W$ be the Walsh--Hadamard matrix 
\begin{equation}
\label{eq:2-Walsh-Hadamard}
W=\frac{1}{\sqrt{2}}
\left(
  \begin{array}{cc}
    1& 1 \\
    1& -1
  \end{array}
\right)
=W^{-1}\ , 
\end{equation}
then we can diagonalize $\sigma_{1}$ as 
$
\sigma_{1}=W\sigma_{3}W^{-1}
$
 by making use of this $W$.

Let us consider an atom with $2$ energy levels $E_{0}$ and $E_{1}$. Its 
Hamiltonian is in the diagonal form given as 
\begin{equation}
H_{0}=
\left(
  \begin{array}{cc}
    E_{0}& 0 \\
    0& E_{1}
  \end{array}
\right).
\end{equation}
This is rewritten as 
\begin{equation}
\label{eq:2}
H_{0}=
\frac{E_{0}+E_{1}}{2}
\left(
  \begin{array}{cc}
    1& 0 \\
    0& 1
  \end{array}
\right)+
\frac{E_{0}-E_{1}}{2}
\left(
  \begin{array}{cc}
    1& 0 \\
    0& -1
  \end{array}
\right)
=\Delta_{0}{\bf 1}_{2}+\Delta_{1}\sigma_{3},
\end{equation}
where $\Delta_{0}=(1/2)(E_{0}+E_{1})$ and $\Delta_{1}=(1/2)(E_{0}-E_{1})$. 
Since we usually take no interest in constant terms, we set 
\begin{equation}
\label{eq:2-hamiltonian}
H_{0}=\Delta_{1}\sigma_{3},\quad \mbox{where}\quad 
\Delta_{1}=(1/2)(E_{0}-E_{1}). 
\end{equation}

Next we would like to extend $2$--level system to general $n$--level system. 
For that we consider the $3$--level one in detail as a test case. 
The Hamiltonian with three energy levels $E_{0}$, $E_{1}$, $E_{2}$ is 
\begin{equation}
H_{0}=
\left(
  \begin{array}{ccc}
    E_{0}& &  \\
    & E_{1}&  \\
    &  & E_{2} 
  \end{array}
\right).
\end{equation}
We want to rewrite this as (\ref{eq:2}). Let $\sigma$ be 
$\mbox{exp}(\frac{2\pi i}{3})$, then it is easy to see 
\begin{equation}
\label{eq:simple-relations}
\sigma^{3}=1,\quad \bar{\sigma}=\sigma^{2},\quad 1+\sigma+\sigma^{2}=0.
\end{equation}
Since the matrix corresponding to $\sigma_{3}$ is considered as 
\begin{equation}
\Sigma_{3}=
\left(
  \begin{array}{ccc}
    1& &       \\
    & \sigma&  \\
    &  & \sigma^{2}
  \end{array}
\right), 
\end{equation}
we solve the equation 
\begin{equation}
H_{0}=\Delta_{0}{\bf 1}_{3}+\Delta_{1}\Sigma_{3}+
\Delta_{2}\Sigma_{3}^{2} 
\end{equation}
with indeterminants $\Delta_{0}$, $\Delta_{1}$, $\Delta_{2}$. 
Then we have 
\begin{equation}
\left\{
\begin{array}{rl}
\Delta_{0}+\Delta_{1}+\Delta_{2}=E_{0} \\
\Delta_{0}+\sigma\Delta_{1}+\sigma^{2}\Delta_{2}=E_{1} \\
\Delta_{0}+\sigma^{2}\Delta_{1}+\sigma\Delta_{2}=E_{2} 
\end{array}
\right.
\quad \Longleftrightarrow \quad 
\left(
  \begin{array}{ccc}
    1& 1& 1      \\
    1& \sigma& \sigma^{2}  \\
    1& \sigma^{2}& \sigma
  \end{array}
\right)
\left(
  \begin{array}{c}
    \Delta_{0} \\
    \Delta_{1} \\
    \Delta_{2} 
  \end{array}
\right)
=
\left(
  \begin{array}{c}
    E_{0} \\
    E_{1} \\
    E_{2} 
  \end{array}
\right),
\end{equation}
so we obtain 
\begin{equation}
\label{eq:3-generalized equation}
\left(
  \begin{array}{c}
    \Delta_{0} \\
    \Delta_{1} \\
    \Delta_{2} 
  \end{array}
\right)
=
\left(
  \begin{array}{ccc}
    1& 1& 1      \\
    1& \sigma& \sigma^{2}  \\
    1& \sigma^{2}& \sigma
  \end{array}
\right)^{-1}
\left(
  \begin{array}{c}
    E_{0} \\
    E_{1} \\
    E_{2} 
  \end{array}
\right)
=\frac{1}{3}
\left(
  \begin{array}{ccc}
    1& 1& 1      \\
    1& \sigma^{2}& \sigma  \\
    1& \sigma& \sigma^{2}
  \end{array}
\right)
\left(
  \begin{array}{c}
    E_{0} \\
    E_{1} \\
    E_{2} 
  \end{array}
\right), 
\end{equation}
or 
\begin{equation}
\left\{
\begin{array}{l}
\Delta_{0}=\frac{1}{3}(E_{0}+E_{1}+E_{2}), \\
\Delta_{1}=\frac{1}{3}(E_{0}+\sigma^{2}E_{1}+\sigma E_{2}), \\
\Delta_{2}=\frac{1}{3}(E_{0}+\sigma E_{1}+\sigma^{2}E_{2}). 
\end{array}
\right.
\end{equation}
%
%
Therefore the Hamiltonian which neglects the constant term is written as 
\begin{equation}
\label{eq:3-hamiltonian}
H_{0}=\Delta_{1}\Sigma_{3}+\Delta_{2}\Sigma_{3}^{2} 
     =\Delta_{1}\Sigma_{3}+\bar{\Delta}_{1}\Sigma_{3}^{\dagger}\ .
\end{equation}

\par \noindent 
These $\Delta_{1}$ and $\Delta_{2}$ may be considered as a complex 
"energy difference" among $E_{0}$, $E_{1}$, $E_{2}$. 

By the way, the matrix in the right hand side of 
(\ref{eq:3-generalized equation}) (replaced $1/3$ with $1/\sqrt{3}$) 
\begin{equation}
\label{eq:3-Walsh-Hadamard}
W=\frac{1}{\sqrt{3}}
\left(
  \begin{array}{ccc}
    1& 1& 1      \\
    1& \sigma^{2}& \sigma  \\
    1& \sigma& \sigma^{2}
  \end{array}
\right)\quad \in \quad U(3)
\end{equation}
is very familiar to us. Let $\Sigma_{1}$ and $\Sigma_{3}$ be generators 
of generalized Pauli matrices in the case of $n=3$, namely 
\begin{equation}
\Sigma_{1}=
\left(
  \begin{array}{ccc}
    0& & 1    \\
    1& 0&      \\
     & 1& 0
  \end{array}
\right), \quad 
\Sigma_{3}=
\left(
  \begin{array}{ccc}
    1& &       \\
    & \sigma&  \\
    &  & \sigma^{2}
  \end{array}
\right).
\end{equation}
Then it is easy to see 
\begin{equation}
\Sigma_{1}^{3}=\Sigma_{3}^{3}={\bf 1}_{3},\quad 
\Sigma_{1}^{\dagger}=\Sigma_{1}^{2}, \quad 
\Sigma_{3}^{\dagger}=\Sigma_{3}^{2}, \quad 
\Sigma_{3}\Sigma_{1}=\sigma \Sigma_{1}\Sigma_{3}.
\end{equation}

Now we can show that 
$\Sigma_{1}$ can be diagonalized by making use of $W$ above 
\begin{equation}
\label{eq:3-diagonalization}
\Sigma_{1}=W\Sigma_{3}W^{\dagger}=W\Sigma_{3}W^{-1}.
\end{equation}
In fact 
\begin{eqnarray}
W\Sigma_{3}W^{\dagger}&=&\frac{1}{3}
\left(
\begin{array}{ccc}
1&          1&     1     \\
1& \sigma^{2}& \sigma    \\
1& \sigma    & \sigma^{2}
\end{array}
\right)
\left(
\begin{array}{ccc}
1&       &           \\
 & \sigma&           \\
 &       & {\sigma}^2 
\end{array}
\right)
\left(
\begin{array}{ccc}
1&        1&     1 \\
1& \sigma & \sigma^{2} \\
1& \sigma^{2}  & \sigma
\end{array}
\right)   \nonumber \\
&=&\frac{1}{3}
\left(
\begin{array}{ccc}
1& \sigma& \sigma^{2}\\
1&      1&  1        \\
1& \sigma^{2}& \sigma 
\end{array}
\right)
\left(
\begin{array}{ccc}
1&        1&     1 \\
1& \sigma& \sigma^{2} \\
1& \sigma^{2}  & \sigma
\end{array}
\right)  
=\frac{1}{3}
\left(
\begin{array}{ccc}
0&  0& 3\\
3&  0& 0\\
0&  3& 0
\end{array}
\right) 
=\Sigma_{1},  \nonumber
\end{eqnarray}
where we have used the relations of $\sigma$ in (\ref{eq:simple-relations}). 

\par \noindent
A comment is in order. Since $W$ corresponds to the Walsh--Hadamard matrix
(\ref{eq:2-Walsh-Hadamard}), so it may be possible to call $W$ 
the generalized Walsh--Hadamard matrix.

Let us proceed with general case. 
The Hamiltonian with $n$ energy levels $E_{0}$, $E_{1}$, $E_{2}$, $\cdots$, 
$E_{n-1}$ is 
\begin{equation}
H_{0}=
\left(
  \begin{array}{cccccc}
    E_{0}& &  & & &  \\
    & E_{1}&  & & &  \\
    &  & E_{2}& & &  \\
    &  & & \cdot& &  \\
    &  & & & \cdot&  \\
    &  &  &  & & E_{n-1}
  \end{array}
\right).
\end{equation}

From the discussion in the case of $n=3$ it is easy to rewrite the 
Hamiltonian. 
Let $\{\Sigma_{1}, \Sigma_{3}\}$ be generalized Pauli matrices (see 
\cite{KF1}, Appendix B)
\begin{equation}
\label{Sigma-1}
\Sigma_{1}=
\left(
\begin{array}{cccccc}
0&  &  &      &      & 1  \\
1& 0&  &      &      &    \\
 & 1& 0&      &      &    \\
 &  & \cdot& \cdot&      &    \\
 &  &  & \cdot& \cdot&    \\
 &  &  &      &      1 & 0
\end{array}
\right),      \qquad 
\label{Sigma-3}
\Sigma_{3}=
\left(
\begin{array}{cccccc}
1&       &           &      &      &             \\
 & \sigma&           &      &      &             \\
 &       & {\sigma}^2&      &      &             \\
 &       &           & \cdot&      &             \\
 &       &           &      & \cdot&             \\
 &       &           &      &      & {\sigma}^{n-1}
\end{array}
\right)
\end{equation}
where $\sigma$ is a primitive element $\sigma=\mbox{exp}(\frac{2\pi i}{n})$ 
which satisfies 
\begin{equation}
\sigma^{n}=1,\quad \bar{\sigma}=\sigma^{n-1},\quad 
1+\sigma+\cdots +\sigma^{n-1}=0.
\end{equation}

\par \noindent
Then it is easy to see 
\begin{equation}
\Sigma_{1}^{n}=\Sigma_{3}^{n}={\bf 1}_{n},\quad 
\Sigma_{1}^{\dagger}=\Sigma_{1}^{n-1}, \quad 
\Sigma_{3}^{\dagger}=\Sigma_{3}^{n-1}, \quad 
\Sigma_{3}\Sigma_{1}=\sigma \Sigma_{1}\Sigma_{3}.
\end{equation}

If we define a Vandermonde matrix $W$ based on $\sigma$ as 
\begin{equation}
\label{eq:n-Walsh-Hadamard}
W=\frac{1}{\sqrt{n}}
\left(
\begin{array}{ccccccc}
1&        1&     1&   \cdot & \cdot  & \cdot & 1             \\
1& \sigma^{n-1}& \sigma^{2(n-1)}&  \cdot& \cdot& \cdot& \sigma^{(n-1)^2} \\
1& \sigma^{n-2}& \sigma^{2(n-2)}&  \cdot& \cdot& \cdot& \sigma^{(n-1)(n-2)} \\
\cdot&  \cdot &  \cdot  &     &      &      & \cdot  \\
\cdot&  \cdot  & \cdot &      &      &      &  \cdot  \\
1& \sigma^{2}& \sigma^{4}& \cdot & \cdot & \cdot & \sigma^{2(n-1)} \\
1& \sigma & \sigma^{2}& \cdot& \cdot& \cdot& \sigma^{n-1}
\end{array}
\right), 
\end{equation}
then it is not difficult to see 
\begin{equation}
\label{eq:n-diagonalization}
\Sigma_{1}=W\Sigma_{3}W^{\dagger}=W\Sigma_{3}W^{-1}.
\end{equation}

\par \noindent 
That is, $\Sigma_{1}$ can be diagonalized by making use of $W$.

We set for simplicity 
\begin{equation}
{\bf \Delta}=(\Delta_{0},\Delta_{1},\cdots,\Delta_{n-2},\Delta_{n-1})^{t}, 
\quad 
{\bf E}=(E_{0},E_{1},\cdots,E_{n-2},E_{n-1})^{t},
\end{equation}
and the relation like (\ref{eq:3-generalized equation}) between them is given 
by 
\begin{equation}
{\bf \Delta}=\frac{1}{\sqrt{n}}W{\bf E}, 
\end{equation}
or explicitly 
\[
\Delta_{j}=(1/n)\left(
E_{0}+\sigma^{n-j}E_{1}+\sigma^{2(n-j)}E_{2}+\cdots+
\sigma^{(n-1)(n-j)}E_{n-1}
\right) 
\]
for\ \ $0\leq j \leq n-1$. Then the Hamiltonian which neglects the constant 
term is written as 
\begin{equation}
\label{eq:n-hamiltonian}
H_{0}=\sum_{j=1}^{n-1}\Delta_{j}\Sigma_{3}^{j}\ . 
\end{equation}

\par \vspace{5mm}
Next let us incorporate an interaction into the model. 
We consider an atom with two energy levels which interacts with external 
(periodic) field with $g\mbox{cos}(\omega t)$. 
The Hamiltonian in the dipole approximation is given by 
\begin{equation}
\label{eq:2-full-hamiltonian}
H=H_{0}+g\ \mbox{cos}(\omega t)\sigma_{1}
=\Delta_{1}\sigma_{3}+g\ \mbox{cos}(\omega t)\sigma_{1}, 
\quad \mbox{cos}(\omega t)=(1/2)(\mbox{e}^{i\omega t}+\mbox{e}^{-i\omega t}), 
\end{equation}
where $\Delta_{1}$ is the constant in (\ref{eq:2-hamiltonian}), 
$\omega$ the frequency of the external field, 
$g$ the coupling constant between the external field and the atom. 
In the following we cannot assume the rotating wave approximation 
(which neglects the fast oscillating terms), namely the Hamiltonian given by 
\begin{equation}
\label{eq:jaynes-cummings}
H=\Delta_{1}\sigma_{3}+\frac{g}{2}\left(\mbox{e}^{i\omega t}\sigma_{+}+
\mbox{e}^{-i\omega t}\sigma_{-}\right). 
\end{equation}

We would like to extend this to the general case. We again note that 
$\Sigma_{1}$ is not hermitian, so we must take this fact into consideration. 
The interaction term corresponding to $\mbox{cos}(\omega t)\sigma_{1}$ is 
considered to be 
\begin{equation}
\frac{1}{2}
\left(\mbox{e}^{i\omega t}\Sigma_{1}+\mbox{e}^{-i\omega t}\Sigma_{1}^{\dagger}
\right) 
\end{equation}
near to the above (\ref{eq:jaynes-cummings}). Let us explain the reason. 
The diagonalized form of $\mbox{cos}(\omega t)\sigma_{1}$ is given by 
\[
\mbox{cos}(\omega t)\sigma_{1}
=W
\left(
  \begin{array}{cc}
    \mbox{cos}(\omega t)&  \\
    & -\mbox{cos}(\omega t)
  \end{array}
\right)
W^{-1}
=W
\left(
  \begin{array}{cc}
    \mbox{cos}(\omega t)&  \\
    & \mbox{cos}(\omega t+\pi)
  \end{array}
\right)
W^{-1}
\]
by making use of $W$ in (\ref{eq:2-Walsh-Hadamard}). 
On the other hand it is easy to see 
\begin{equation}
\frac{1}{2}
\left(\mbox{e}^{i\omega t}\Sigma_{1}+\mbox{e}^{-i\omega t}\Sigma_{1}^{\dagger}
\right)
=WDW^{-1}
\end{equation}
by making use of $W$ in (\ref{eq:n-Walsh-Hadamard}), where D is a diagonal 
matrix 
\begin{equation}
D=
\left(
\begin{array}{cccccc}
\mbox{cos}(\omega t)&       &           &      &      &     \\
 & \mbox{cos}(\omega t+2\pi/n)&           &      &      &   \\
 &       & \cdot&      &      &       \\
 &       &           & \cdot&      &                        \\
 &       &           &      & \cdot&                        \\
 &       &           &      &      & \mbox{cos}(\omega t+2\pi(n-1)/n)
\end{array}
\right). 
\end{equation}
Therefore this is a natural generalization of 
$\mbox{cos}(\omega t)\sigma_{1}$.

Now we are in a position to state our proposal : 

\par \noindent 
{\bf Proposal}\quad The Hamiltonian in the general case is 
\begin{equation}
\label{eq:n-full-hamiltonian}
H=H_{0}+
\frac{g}{2}
\left(\mbox{e}^{i\omega t}\Sigma_{1}+\mbox{e}^{-i\omega t}\Sigma_{1}^{\dagger}
\right)
=\sum_{j=1}^{n-1}\Delta_{j}\Sigma_{3}^{j}+
\frac{g}{2}
\left(\mbox{e}^{i\omega t}\Sigma_{1}+\mbox{e}^{-i\omega t}\Sigma_{1}^{\dagger}
\right) 
\end{equation}
where $g$ is a coupling constant of the system. 
In particular in the case of $n=3$ 
\begin{equation}
H=\Delta_{1}\Sigma_{3}+\bar{\Delta}_{1}\Sigma_{3}^{\dagger}+
\frac{g}{2}
\left(\mbox{e}^{i\omega t}\Sigma_{1}+\mbox{e}^{-i\omega t}\Sigma_{1}^{\dagger}
\right) 
\end{equation}
by (\ref{eq:3-hamiltonian}). 

We have written down the Hamiltonian in the general case in terms of 
generalized Pauli matrices, however don't know whether or not it is actually 
reasonable in Quantum Optics or Condensed Matter Physics (refer to 
\cite{MSIII}). 
Further considerations will be needed. 

We conclude this paper by stating that constructions of solutions under 
some ansatz (in the weak or strong coupling regime) are not difficult and 
will be published in another paper, \cite{KF6}.


\end{document}